# Large in-plane optical anisotropy in epitaxial CrN(113) films

Duc V. Dinh,[*,†] Frank Peiris,[‡] and Oliver Brandt[†]

†Paul-Drude-Institut für Festkörperelektronik, Leibniz-Institut im Forschungsverbund Berlin e.V., Hausvogteiplatz 5–7, 10117 Berlin, Germany
‡Department of Physics, Kenyon College, Gambier, 43022 Ohio, United States

E-mail: duc.vandinh@pdi-berlin.de

## Abstract

**CrN, a transition-metal nitride crystallizing in the cubic rocksalt structure, exhibits a unique combination of mechanical robustness, semiconducting behavior, and thermoelectric functionality. To date, however, CrN has been synthesized primarily as polycrystalline material or as epitaxial films grown along high-symmetry orientations such as (001) and (111). Here, we prepare epitaxial CrN(113) films on AlN(11$\bar{2}$2) templates by plasma-assisted molecular beam epitaxy. The low-symmetry interface and highly direction-dependent lattice mismatch are expected to modify the physical properties of CrN relative to the bulk material. Indeed, variable-angle spectroscopic ellipsometry in the photon-energy range from 1.25 to 5 eV reveals a surprisingly strong polarization dependence of the optical response despite the nearly strain-free cubic crystal structure of the films. The pseudo-optical constants *n* and *k* differ by more than 10% for orthogonal in-plane polarizations throughout the investigated spectral range. We evaluate possible origins of this effect and conclude that it most likely arises from anisotropic surface morphology of CrN(113), which creates a metasurface acting as a subwavelength grating in the strongly absorbing regime.**

## 1 Introduction

Together with ScN, CrN forms a distinct class of transition-metal nitrides that crystallize in the cubic rocksalt structure, possess polar bonds giving rise to a pronounced Reststrahlen band, and are indirect semiconductors.[1–3] CrN additionally combines mechanical robustness—including high hardness, wear resistance, and excellent corrosion stability[4,5]—with electronic,[2,6–11] magnetic,[2,12–16] and thermoelectric functionalities,[17–21] reflecting its potential for coupled spin, charge, and heat transport phenomena.

Following the establishment of its semiconducting nature in recent work,[2,3] a systematic study of the optical properties of CrN is warranted. Spectroscopic ellipsometry (SE) is widely used to determine the complex dielectric function of transition-metal nitrides, including ScN,[1,22,23] TiN,[24,25] ZrN,[26] as well as CrN.[3,27] We recently extended this approach to epitaxial, semiconducting CrN(111) layers, identifying two direct band gaps[3] in addition to a fundamental indirect gap detected electrically.[2] Consistent with the threefold sym-metry of the (111) surface, the in-plane optical response was isotropic, as expected for cubic materials.

However, even in cubic crystals, optical anisotropy can arise from second order effects related to intrinsic and extrinsic mechanisms in the bulk or, particularly, at surfaces. Such effects have been extensively studied for low-symmetry surfaces of semiconductors and oxides, where inequivalent in-plane crystallographic directions lead to anisotropic optical responses, as demonstrated for Si(110), Ge(110), GaAs(110), GaP(110), Si(113), Ge(113), $SrTiO_3$(110), and $TiO_2$(110).[28–34] In contrast, analogous studies on transition-metal nitrides remain scarce.

Here, we investigate epitaxial CrN(113) layers grown on AlN(11$\bar{2}$2) templates by plasma-assisted molecular beam epitaxy. X-ray diffractometry is used to determine crystallographic orientation, epitaxial relationships, and strain state, while variable-angle spectroscopic ellipsometry provides polarization-resolved optical constants over the photon-energy range 1.25–5.0 eV. The combination of reduced surface symmetry and highly anisotropic lattice mismatch provides a platform for assessing orientation-dependent optical response in cubic CrN.

## 2 Experimental

Intentionally undoped 200-nm-thick CrN(113) layers are grown by PAMBE on semipolar AlN(11$\bar{2}$2) templates, which were prepared on $Al_2O_3$(10$\bar{1}$0) substrates.[35,36] The growth conditions are identical to those used for CrN(111) layers and have been reported elsewhere.[2,3]

The structural properties of the layers are characterized using a high-resolution (HR)XRD system (Philips Panalytical X'Pert PRO MRD) equipped with a two-bounce hybrid monochromator Ge(220) for the $CuK_{\alpha 1}$ source ($\lambda$ = 1.540 598 Å). The surface morphology of layers is imaged by atomic force microscopy in contact mode (Dimension Edge, Bruker).

SE measurements were carried out at room temperature over the spectral range of 1.25–5.0 eV using a SOPRA GES-5E ellipsometer. Before acquiring data on the CrN(113) layers, the ellipsometer performance was validated using a reference ≈ 125-nm-thick $SiO_2$/Si layer. SE data were collected at four angles of incidence (AOI) $\Theta$, namely, 55°, 60°, 65°, and 70° with respect to the surface normal. The measured ellipsometric spectra of this reference sample were fitted using a standard Cauchy model supplemented with a weak sub-gap absorption term (band-tails), which is required to accurately reproduce the small but finite absorption of $SiO_2$ and to obtain a high-quality fit. The fit yields an oxide thickness in excellent agreement with the nominal thickness value (see Fig. S1 in the Supplemental Material). This result confirms correct instrument calibration and alignment, ensuring that the measurements with similar AOIs from the CrN(113) samples are not influenced by systematic errors. The optical constants of the CrN layers are extracted by fitting the SE data using a parametric multilayer model described in more detail below.

## 3 Results and Discussion

### 3.1 Structural properties

To investigate the epitaxial orientation of the CrN layers grown on AlN(11$\bar{2}$2) templates, symmetric $2\theta - \omega$ XRD scans are performed over a wide angular range with open detector. Figure 1(a) shows such XRD scans along the $[1\bar{1}00]_{AlN}$ direction for a 200-nm-thick CrN layer. Initially, measurements aligned on the $Al_2O_3$ 30$\bar{3}$0 reflection with negligible tilt (i.e., $\psi = 0°$) only show the $Al_2O_3$ 30$\bar{3}$0 and AlN 11$\bar{2}$2 reflections, indicating that these surface planes are exactly parallel with respect to each other. By tilting the sample to $\psi \approx 6°$, the $Al_2O_3$ and AlN reflections could no longer be detected, but instead a pronounced reflection at about 76° became apparent, which we attribute to the CrN(113) reflection [calculated based on the relaxed lattice constant $a_0$ = 4.145 Å measured for CrN(111) layers[2]].

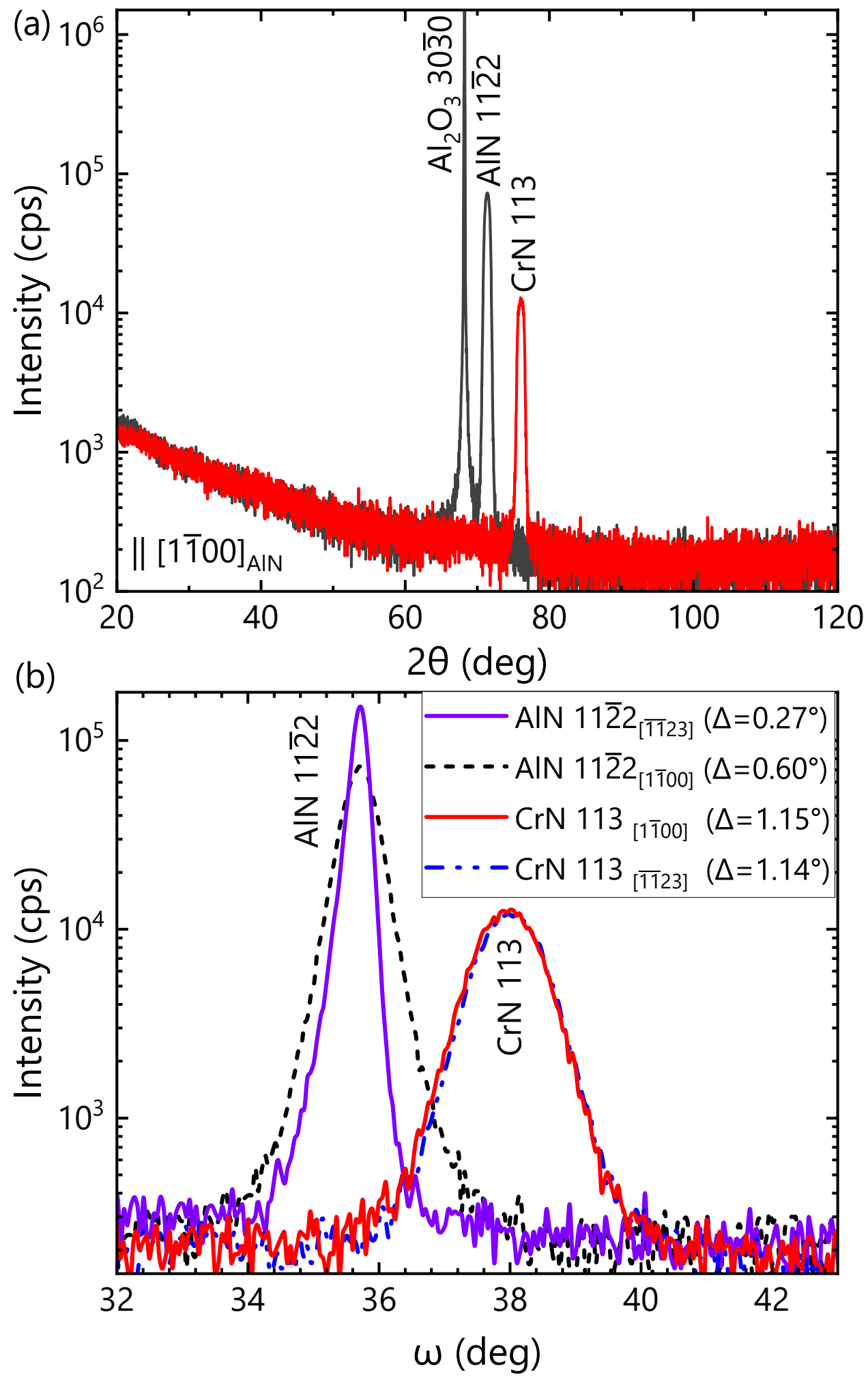

Figure 1: (a) Symmetric $2\theta-\omega$ XRD scans of the 200-nm-thick CrN layer grown on $AlN(11\bar{2}2)/Al_2O_3(10\bar{1}0)$ template performed along the $[1\bar{1}00]_{AlN}$ direction with $\psi = 0°$ and $\psi = 6°$. (b) XRCs across the $AlN11\bar{2}2$ and CrN113 reflections along the $[\bar{1}\bar{1}23]_{AlN}$ and $[1\bar{1}00]_{AlN}$ directions. All these measurements were performed with an open detector without any receiving slit.

This finding is explained by symmetric reciprocal space maps (RSM) recorded along the $[\bar{1}\bar{1}23]_{AlN}$ and $[1\bar{1}00]_{AlN}$ directions, as shown in Fig. 2. The map measured along the $[\bar{1}\bar{1}23]_{AlN}$ direction shows that the $AlN11\bar{2}2$ reflection is already tilted by approximately $\omega \approx 1.7°$ relative to the $Al_2O_3 30\bar{3}0$ reflection, while the CrN113 reflection exhibits an additional $\omega \approx 3.5°$ tilt with respect to the AlN reflection [Fig. 2(a)]. This $\omega$-shift reflects a rotation of the CrN lattice normal within the diffraction plane, corresponding to a rotation about an in-plane axis parallel to $[1\bar{1}0]_{CrN} \parallel [\bar{1}\bar{1}23]_{AlN}$ (see atomic structures in Fig. 3). In contrast, when probing along the orthogonal in-plane $[33\bar{2}]_{CrN} \parallel [1\bar{1}00]_{AlN}$ direction, the map acquired at $\psi = 0°$ reveals only the $Al_2O_3$ and AlN reflections, indicating that no additional $\omega$-tilt is present along this azimuth. The CrN113 reflection appears only when a $\psi$-tilt of $\approx 6°$ is applied, demonstrating that the CrN unit cell is also rotated out of the diffraction plane – i.e., about an in-plane axis parallel to the $[1\bar{1}0]_{CrN}$ direction.

Because the lattice mismatch between CrN and AlN is strongly anisotropic along the $[1\bar{1}0]_{CrN}$ and $[33\bar{2}]_{CrN}$ directions, the resulting strain state can slightly modify the (113) interplanar spacing in a direction-dependent manner, producing a small azimuth-dependent shift in $Q_z$ for the CrN(113) reflection. Together, the $\omega$- and $\psi$-offsets show that the CrN layer possesses a two-component lattice tilt within the (113) plane, arising from rotations about in-plane axes parallel to the $[33\bar{2}]_{CrN}$ and $[1\bar{1}0]_{CrN}$ directions. These tilts modify the projection of the (113) reciprocal-lattice vector onto $Q_z$ measured along the two in-plane directions, producing small azimuth-dependent shifts that are geometric rather than intrinsic changes in lattice spacing. This behavior is consistent with the strongly anisotropic lattice mismatch between CrN and AlN – approximately −0.14% along the $[1\bar{1}0]_{CrN}$ direction and −9.82% along the $[33\bar{2}]_{CrN}$ direction – which initially generates a slightly biaxial tensile strain state (Fig. 3). Although the CrN(113) layer is fully relaxed in the final state, exhibiting an out-of-plane lattice constant of 4.146 Å and in-plane lattice constants of 2.933 Å along $[1\bar{1}0]_{CrN}$ and 9.727 Å along $[33\bar{2}]_{CrN}$ (see Fig. S2 in the Supplemental Material), the strongly anisotropic mismatch during the early stages of epitaxy likely drives direction-dependent strain accommodation, leading to the observed in-plane lattice tilt and azimuth-dependent reciprocal-space shifts that remain as structural signatures after relaxation.

Both X-ray rocking curves [Fig. 1(b)] and RSM measurements [Figs. 2(a)-2(b)] clearly show anisotropic broadening of the AlN11$\bar{2}$2 reflection along the two orthogonal in-plane directions. Interesting, the broadening of the CrN113 rocking curves along these directions are comparable, and about $2-3\times$ smaller than that of 75-nm-thick sample reported in Ref. 2. This reduction in broadening is likely associated with the larger layer thickness in the present sample, which promotes dislocation bending and partial annihilation during growth, thereby reducing the overall mosaic spread.[37–39]

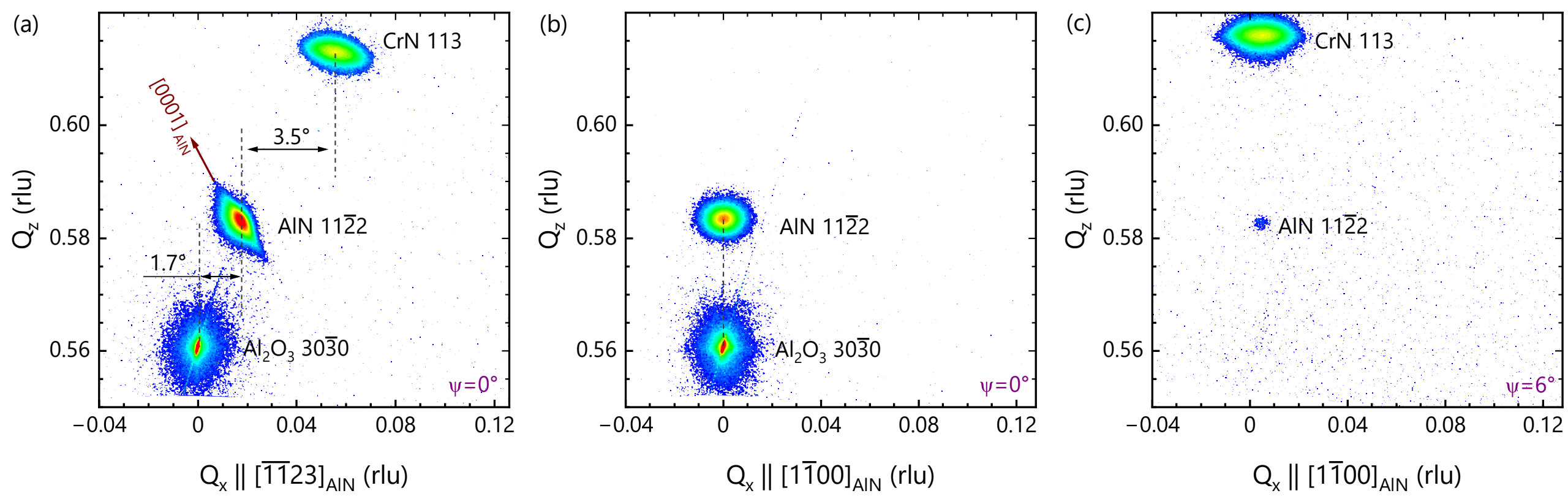

Figure 2: Reciprocal space maps of the 200-nm-thick CrN(113) layer grown on AlN(11$\bar{2}$2) template measured along (a) the [$\bar{1}\bar{1}$23]$_{\text{AlN}}$ ($\psi = 0°$), (b) [1$\bar{1}$00]$_{\text{AlN}}$ ($\psi = 0°$), and (c) [1$\bar{1}$00]$_{\text{AlN}}$ ($\psi = 6°$) directions. These measurements were performed using a 1-mm-wide receiving slit. In (a), the broadening along the [0001]$_{\text{AlN}}$ direction is due to basal-plane stacking faults.[35,36] The diagonal line across the $Al_2O_3$30$\bar{3}$0 reflections are due to an instrumental artifact, namely, the detector streak.

## 3.2 Surface morphology

Figure 4(a) shows the surface morphology of the CrN(113) layer grown on the AlN(11$\bar{2}$2) template. Similar to ScN(113),[40] the CrN(113) film replicates the anisotropic, stripe-like morphology of the underlying semipolar AlN layer[35,36] [see Fig. S3(a) in the Supplemental Material] but exhibits a slightly larger root-mean-square (rms) roughness, with an rms value of 5.2 nm for the $10 \times 10\,\mu m^2$ scan compared to 3.0 nm for the template [see Fig. S3(b) in the Supplemental Material]. Figure 4(b) shows that this increase in roughness is due to the three-dimensional (3D) growth of CrN under the N*-rich conditions utilized here.[2,3] Despite this 3D growth, the surface retains a highly anisotropic morphology characterized by elongated ridges separated by trenches both running along the [1$\bar{1}$0]$_{\text{CrN}}$ direction. The pronounced uniaxial anisotropy is confirmed by the fast Fourier transform (FFT) pattern shown in Fig. S3(c) in the Supplemental Material. The diffuse nature of the fundamental reflections in this pattern as well as the lack of higher order maxima attest to the irregular spacing of the ridges, i. e., the absence of a well-defined periodicity.

## 3.3 In-plane optical anisotropy

SE performed along two orthogonal azimuths (i.e., $\phi \approx 0°$ and $90°$) provides only a rough estimate of anisotropic optical constants, because such two-angle measurements are highly sensitive to small azimuthal misalignments and do not exploit the full angular dependence of the dielectric response. This limitation has been recognized in anisotropic surface ellipsometry studies,[41–45] where azimuth-dependent measurements are required to reliably resolve in-plane dielectric tensor components. Importantly, the multi-azimuth approach enables a self-consistent reconstruction of the angular dependence of the optical response, allowing the optical constants along the principal axes to be determined directly from the data. This overdetermined fitting strategy significantly improves the robustness and accuracy of the extracted dielectric tensor compared to conventional two-angle measurements.

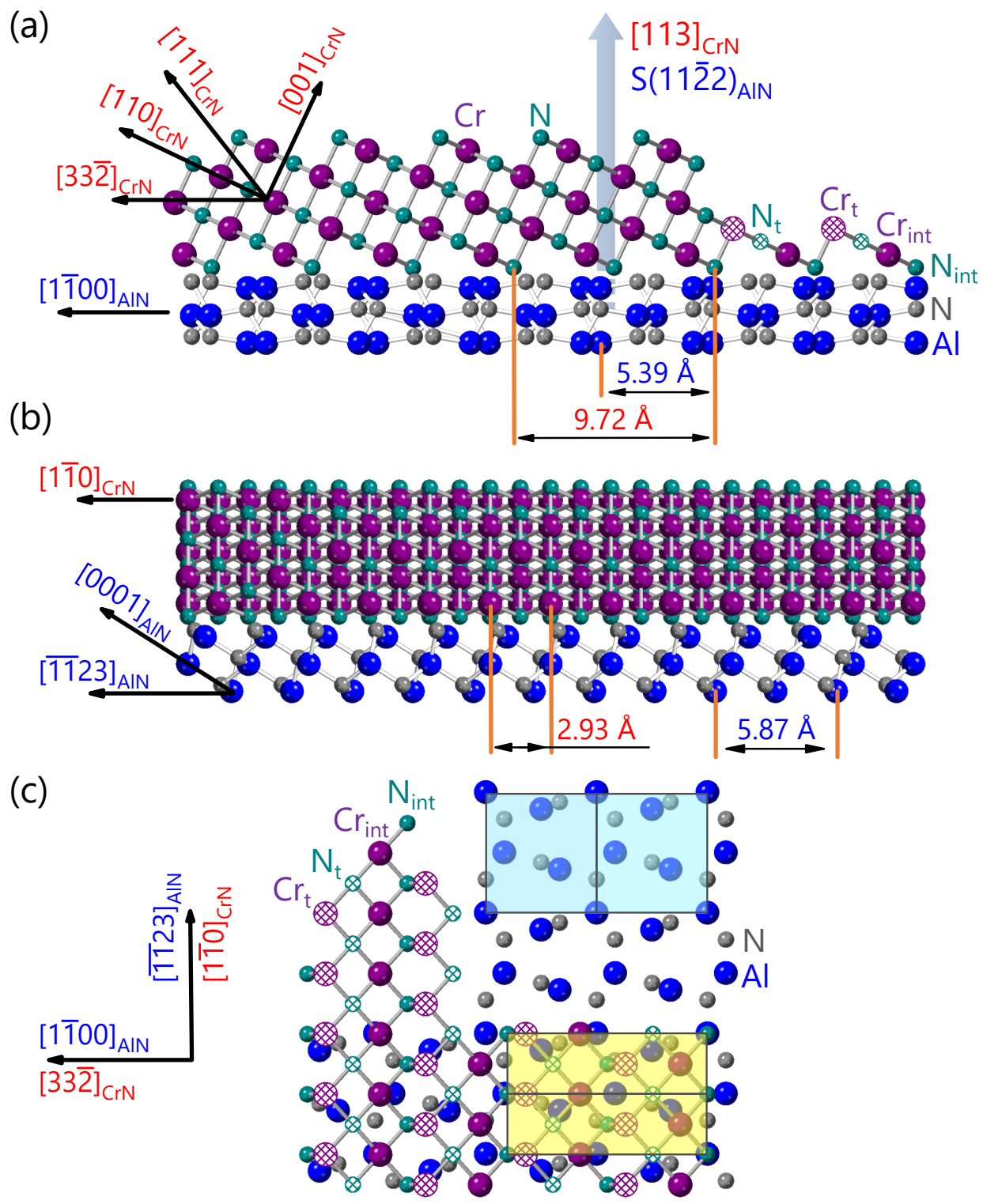

Figure 3: Side views of the atomic structures of CrN(113)/AlN(11$\bar{2}$2): (a) along the $[33\bar{2}]_{\mathrm{CrN}}$ and $[1\bar{1}00]_{\mathrm{AlN}}$ zone axes, and (b) along the $[1\bar{1}0]_{\mathrm{CrN}}$ and $[\bar{1}\bar{1}23]_{\mathrm{AlN}}$ zone axes. $S(11\bar{2}2)_{\mathrm{AlN}}$ denotes the surface normal of the AlN(11$\bar{2}$2) plane. (c) Top view of CrN(113) on the AlN(11$\bar{2}$2) surface. The shaded rectangles visualize the $1 \times 2$ and $2 \times 1$ unit cells of CrN(113) and AlN(11$\bar{2}$2), respectively. In (c), compare the positions of the interfacial and top $Cr_{t,int}$ and $N_{t,int}$ atoms to those depicted in panel (a).

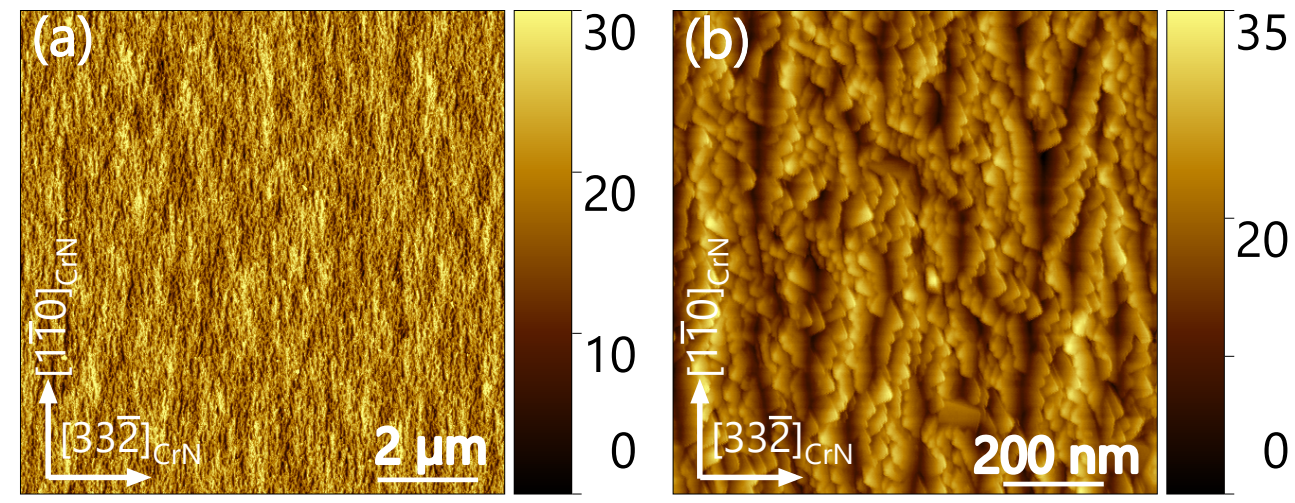

Figure 4: (a) $10 \times 10\,\mu\mathrm{m}^2$ and (b) $1 \times 1\,\mu\mathrm{m}^2$ atomic force topographs of the 200-nm-thick CrN(113) layer. The $z$ scale is in units of nm. Both topographs have a similar root-mean-square surface roughness of 5.2 nm.

Thus, to determine these tensor components on the CrN(113) surface, nine sets of ellipsometric spectra $(\Psi, \Delta)$ were measured at different azimuthal orientations $\phi_j$ ($j = 0, \ldots, 8$) spanning from 0° to 180°. The nominal $\phi_0 \approx 0°$ alignment was initially set close to the $[33\bar{2}]_{\mathrm{CrN}}$ direction based on prior XRD/AFM characterization. A CrN/AlN/$Al_2O_3$ layer structure was used to construct an ellipsometric model. Additionally, surface roughness is accounted for by an effective medium layer at the CrN surface. Note that we did not attempt to model the uniaxial anisotropy of the surface in view of its rather complex morphology [cf. Fig. 4(b)].

All azimuthal datasets were fitted using a Kramers–Kronig-consistent B-spline parameterization, yielding the refractive index $n(\phi_j, E_i)$ and the extinction coefficient $k(\phi_j, E_i)$ at each photon energy $E_i$, as exemplified for $\phi = 0°$ and 90° in Fig. 5, with additional angles shown in Fig. S4 in the Supplemental Material. Despite the measurable topographic roughness, the SE fits did not require an explicit surface-roughness layer, consistent with the fact that a Kramers–Kronig-consistent B-spline absorbs small roughness-related effects into the extracted dielectric function and that the optical response is dominated by the bulk CrN film.

The extracted pseudo-optical constant $n(\phi_j)$ exhibits identical spectral shape for all azimuths and continuously evolves between $n_{\mathrm{min}}$ ($\phi = 0°$) and $n_{\mathrm{max}}$ ($\phi = 90°$) upon rotation, indicating a well-defined principal-axis response and negligible depolarization, as shown in Figs. 6(a)–6(b). The corresponding $\Psi(\phi_j)$ and $\Delta(\phi_j)$ spectra exhibit the same smooth and systematic angular evolution, reflecting the continuous variation of the underlying $n(\phi_j)$ and $k(\phi_j)$ (see Fig. 5 in the Supplemental Material). For the pseudo-optical constant $k(\phi_j)$, the steep rise below 1.7 eV is due to a higher direct interband transition as reported in Ref. 3. The variation of $k$ with different $(\phi_j)$ becomes stronger with increasing photon energy, and $k(\phi_j)$ exhibits a smooth and systematic angular evolution similar to $n(\phi_j)$. For completeness, we note that the trends of $n(\phi_j)$ and $k(\phi_j)$ observed for $\phi = 0 \rightarrow 90°$ reverse for $\phi = 90 \rightarrow 180°$, as expected from the twofold in-plane symmetry of the dielectric tensor. Each azimuthal measurement therefore provides a complex dielectric function

$$\varepsilon_j(\phi_j, E_i) = \left[n(\phi_j, E_i) + i\,k(\phi_j, E_i)\right]^2. \quad (1)$$

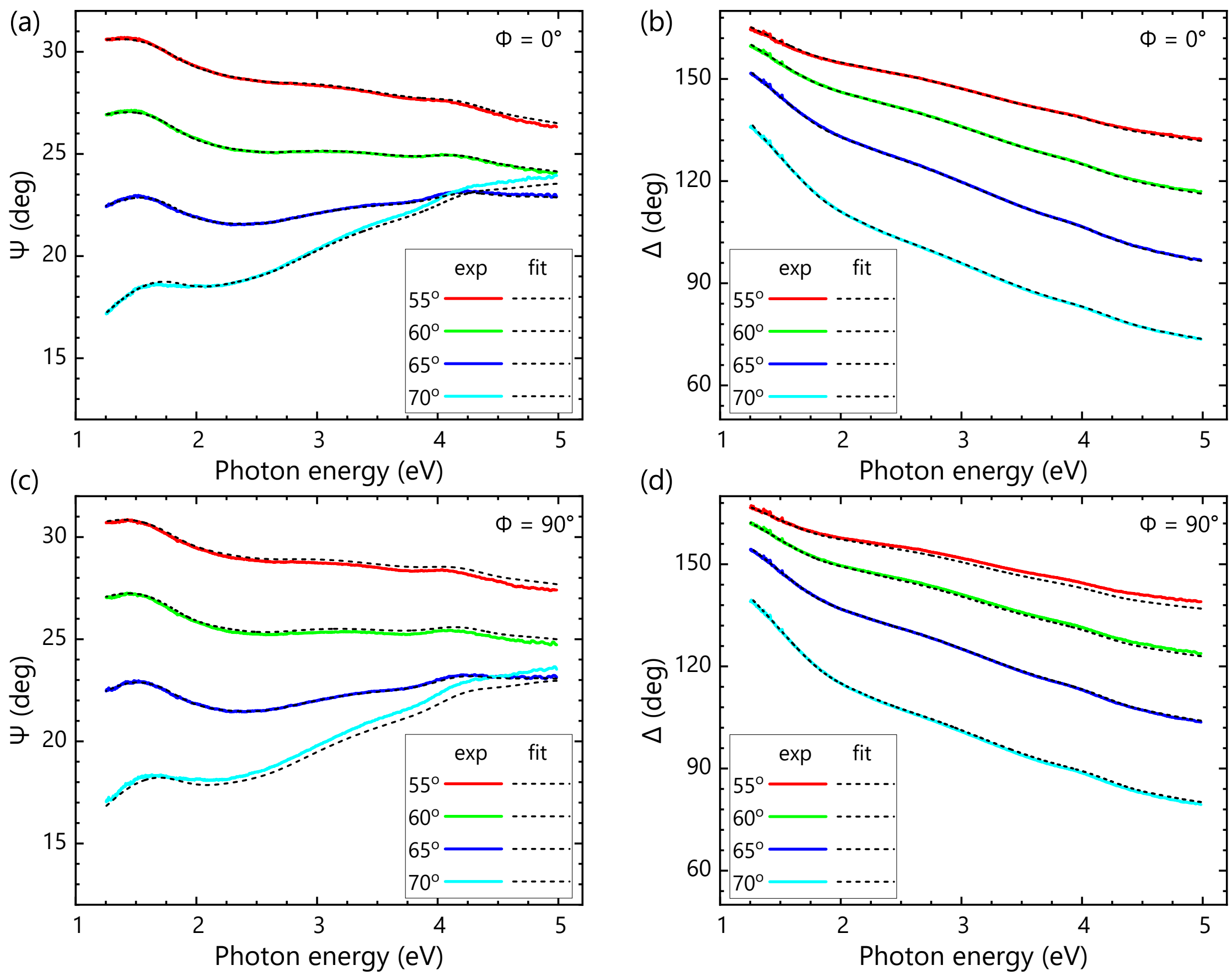

Figure 5: (a) Ψ and (b) Δ ellipsometric spectra measured at $\phi = 0°$ for the CrN(113) film, together with their corresponding fits. Panels (c) and (d) show the analogous data acquired at $\phi = 90°$.

Since the 200-nm-thick CrN(113) layer is strongly absorbing in the entire spectral range between 1.25–5 eV,[3] the contribution of the underlying AlN layer and $Al_2O_3$ substrate to the optical response is negligible, i. e., the measured Ψ and Δ spectra originate almost entirely from CrN. Due to the large refractive index $n$ of CrN, incident light rays strongly bend towards the surface normal even for higher angles of incidence, and consequently decrease the out-of-plane component of $p$-polarized light. Both $p$-polarized and $s$-polarized light thus primarily probes the in-plane dielectric response of the CrN film.[46] In addition, due to the very large extinction coefficient $k$ of CrN within this spectral range,[3] the optical penetration depth is limited to approximately 20–40 nm in the visible and UV spectral range (see Fig. S6 in the Supplemental Material), resulting in a strongly attenuated electromagnetic field inside the film. In this regime, the sensitivity of SE to the out-of-plane dielectric response ($\varepsilon_{zz}$) is reduced, as the measured Ψ and Δ spectra reflect an effective optical response primarily determined by the in-plane dielectric functions and the boundary conditions at the film interfaces.[46,47] Consequently, the influence of $\varepsilon_{zz}$ on the fitted optical response is weak within the experimental sensi-tivity of the present configuration (see Fig. S7 in the Supplemental Material).

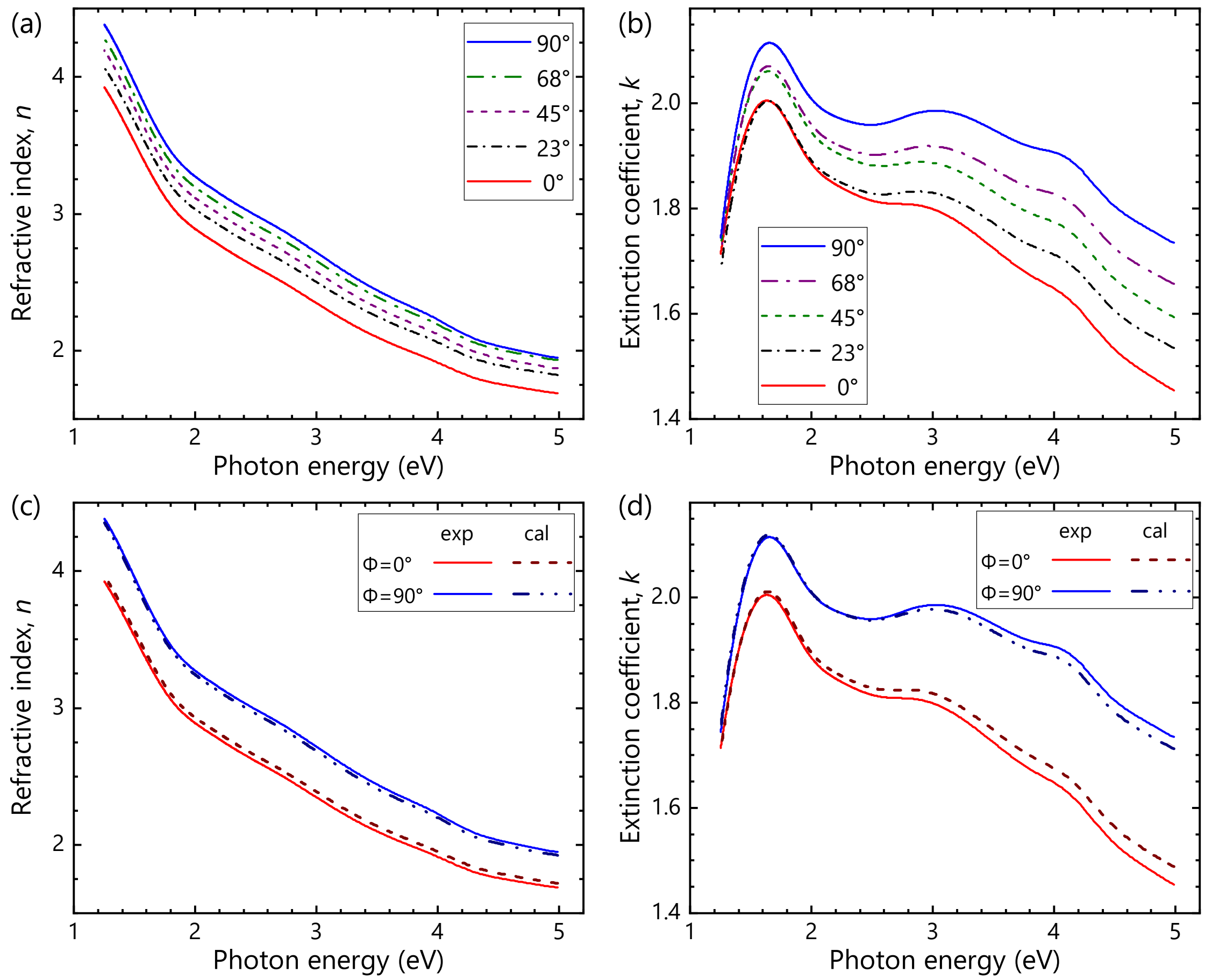

Figure 6: Pseudo-optical constants (a) $n$ and (b) $k$ of the CrN(113) film obtained at different azimuths $\phi_j = 0^\circ \to 90^\circ$. Calculated (c) $n$ and (d) $k$ (dashed lines) for $\phi = 0^\circ$ (along the $[33\bar{2}]_{\mathrm{CrN}}$ direction) and $90^\circ$ (along the $[1\bar{1}0]_{\mathrm{CrN}}$ direction) are shown together with the corresponding experimental data (solid lines) measured at the same azimuthal orientations.

The CrN(113) surface is spanned by the two orthogonal and crystallographically inequivalent in-plane $[33\bar{2}]_{\mathrm{CrN}}$ and $[1\bar{1}0]_{\mathrm{CrN}}$ directions, defining the principal $x$- and $y$-axes. Because of the low sensitivity of the present SE measurements to $\varepsilon_{zz}$ as discussed above, we approximate the optical response of the film by its in-plane dielectric functions, hence neglecting $\varepsilon_{zz}$ altogether. In this basis, the in-plane dielectric tensor is diagonal at each photon energy,[41–45]

$$\varepsilon(E_i) = \begin{pmatrix} \varepsilon_{xx}(E_i) & 0 \\ 0 & \varepsilon_{yy}(E_i) \end{pmatrix}, \quad (2)$$

with four real unknowns: $\Re(\varepsilon_{xx})$, $\Re(\varepsilon_{yy})$, $\Im(\varepsilon_{xx})$, and $\Im(\varepsilon_{yy})$. This diagonal form implies that no polarization mode conversion occurs in the principal-axis frame, allowing the optical response to be fully described within the Jones formalism without requiring a Müller matrix treatment.

In this framework, a small residual misalignment $\gamma$ during sample mounting or rotation can shift the nominal angles such that

$$\phi_j^{\mathrm{tr}} = \phi_j + \gamma. \quad (3)$$

However, acquiring spectra at nine different azimuths provides an overdetermined dataset that enables a robust least-squares reconstruction of $\varepsilon_{xx}(E_i)$, $\varepsilon_{yy}(E_i)$, and $\gamma$. For a diagonal in-plane tensor, the angular dependence is given by

$$\varepsilon(\phi_j^{\mathrm{tr}}, E_i) = \cos^2(\phi_j^{\mathrm{tr}})\, \varepsilon_{xx} + \sin^2(\phi_j^{\mathrm{tr}})\, \varepsilon_{yy}. \quad (4)$$

The nine azimuthal measurements provide nine complex equations (i.e., 18 real constraints), sufficient to determine the five unknown variables. A least-squares solution yields the tensor components $\varepsilon_{xx}(E_i)$, $\varepsilon_{yy}(E_i)$, and the global azimuthal offset $\gamma$.

The principal dielectric responses correspond to the eigenvalues $\eta_{1,2}(E_i)$ of the fitted tensor. Taking the complex square roots of the eigenvalues yields the principal complex refractive indices,

$$n_1(E_i) = \sqrt{\eta_1(E_i)}, \tag{5a}$$

$$n_2(E_i) = \sqrt{\eta_2(E_i)}. \tag{5b}$$

We assign these to the crystallographic directions according to

$$n_{[33\bar{2}]} = \Re(n_1(E_i)), \tag{6a}$$

$$n_{[1\bar{1}0]} = \Re(n_2(E_i)), \tag{6b}$$

with the corresponding extinction coefficients given by

$$k_{[33\bar{2}]} = \Im(n_1(E_i)), \tag{7a}$$

$$k_{[1\bar{1}0]} = \Im(n_2(E_i)). \tag{7b}$$

Figures 6(c)–6(d) show the optical constants calculated from the tensor extracted using the azimuthal datasets and corresponding experimental data for $\phi = 0°$ (along the $[33\bar{2}]_{CrN}$ direction) and $\phi = 90°$ (along the $[1\bar{1}0]_{CrN}$ direction). The azimuthal offset determined by the fits amounts to $\gamma = 2.5°$, indicating that the initial sample alignment was already close to the true $[33\bar{2}]_{CrN}$ crystallographic axis. Note that this angular offset is sufficiently small to yield nearly identical results for calculations setting it to zero; however, for completeness and numerical precision, we retain the offset in all subsequent calculations. These calculated $n(\phi)$ and $k(\phi)$ curves are entirely consistent with the experimental values (see Fig. S5 in the Supplemental Material for a direct comparison of these azimuthal datasets). The small remaining deviation is expected, because each azimuthal spectrum is fitted independently in the SE model, so the resulting $n(\phi)$ and $k(\phi)$ do not follow the exact $\cos^2(\phi)/\sin^2(\phi)$ dependence of a single in-plane dielectric tensor. When these nine slightly different optical constants are combined, the tensor extraction naturally returns the best global match rather than an exact reproduction of every angle. Nevertheless, the excellent agreement between calculated and experimental curves confirms the internal consistency of the measurements, the quality of the SE fitting, and the minimal depolarization in the system.

# 4 Discussion

As already mentioned above, the dielectric tensor of an ideal cubic material is, to first order, a scalar, and the refractive index is thus expected to be represented by a perfectly spherical optical indicatrix in the bulk. However, the optical isotropy of actual cubic crystals can be lifted by several mechanisms, as summarized in Tab. 1. Depending on the mechanism, the resulting in-plane optical anisotropy expressed by the relative change of reflectance $\Delta R/R$ can vary over several orders of magnitude with $\Delta R/R$ ranging from $10^{-4}$ to $10^{-1}$ (i.e., 0.01–10%). Intrinsic mechanisms, regardless of whether they originate in the bulk or at the surface, tend to have comparatively small magnitudes of up to approximately 0.1%. Extrinsic effects can be larger, although the most common ones originating from surface reconstructions[32,45,54,58] and surface electric fields[57,58] typically produce values below 1%. Larger anisotropies of up to $\approx 10\%$ can arise from strain in the bulk and a corresponding loss of cubic symmetry.[50–53] Similarly large anisotropies related to the surface are known from metamaterial and metasurface effects, resulting in form birefringence[59,60] and the effective formation of subwavelength gratings.[61–64]

The low-symmetry CrN(113) layers studied here exhibit a $\Delta R/R$ of 10–12% for $n$ and 3–15% for $k$ along the two principal in-plane directions, exceeding the values associated with intrinsic effects by more than one order of magnitude. Since the layers have been found to be essentially fully relaxed, strain-induced contributions can be ruled out as well. As shown above, however, the CrN(113) layer replicates the anisotropic surface morphology of the AlN(11$\bar{2}$2) template (Fig. 4 and Fig.S3 in the Supplemental Material), exhibiting pronounced striations along the $[\bar{1}\bar{1}23]_{AlN}$ direction ($[1\bar{1}0]_{CrN}$). These features occur on a scale smaller than the wavelength range used for SE measurements (≈ 250–990 nm), i. e., they effectively form a subwavelength grating capable of inducing an in-plane optical anisotropy. A quantitative description of this grating would require full-wave electromagnetic simulations of the actual surface morphology. However, because of the complex surface features of the CrN(113) layer, such simulations are far from trivial and would constitute a study in their own right. To obtain a simplified description suitable for determining the effective optical constants of the grating while retaining the optical isotropy of the underlying CrN layer, the grating can be viewed as an effective metasurface,[63,64] whose optical properties can be approximated using an effective-medium approach.[65,66] The effective medium would be constructed to yield a direction-dependent dielectric response, as commonly done for subwavelength-structured metasurface and metamaterial systems.[61,62] In the present work, we restrict our analysis to the pseudo-optical constants shown in Fig. 6. A more detailed modeling approach based on these considerations will be presented in a forthcoming publication.

Table 1: Classification of mechanisms that can give rise to optical anisotropy in (nominally) cubic crystals. Mechanisms of both intrinsic or extrinsic nature are classified according to the origin of the symmetry breaking mechanism responsible for the anisotropic optical response. In this scheme, a mechanism is designated as surface-related if the anisotropy would disappear upon removal of the surface or interface. Typical magnitudes refer to the relative reflectance anisotropy $\Delta R/R$.

| **Nature** | **Origin** | **Mechanism** | $\Delta R/R$ |
|---|---|---|---|
| Intrinsic | Bulk | Spatial dispersion (Refs. 48,49) | $10^{-4}$–$10^{-3}$ |
| Extrinsic | | Strain (photo-elastic effects) (Refs. 50–53) | $10^{-3}$–$10^{-1}$ |
| Intrinsic | Surface | Many-body screening (Refs. 28,54) | $10^{-4}$–$10^{-3}$ |
| | | Local-field effects (Refs. 55,56) | $10^{-4}$–$10^{-3}$ |
| Extrinsic | | Surface electric fields (electro-optic effects) (Refs. 57,58) | $10^{-4}$–$10^{-2}$ |
| | | Surface reconstruction (Refs. 32,45,54,58) | $10^{-4}$–$10^{-2}$ |
| | | Form birefringence (Refs. 59,60) | $10^{-2}$–$10^{-1}$ |
| | | Sub-wavelength grating effects (Refs. 61–64) | $10^{-1}$ |

To critically examine the interpretation that the optical anisotropy originates from the highly anisotropic surface morphology of the CrN(113) surface, we performed SE measurements on a 200-nm-thick CrN(111)/AlN(0001) reference sample, which was grown side-by-side with the CrN(113) layer. Measurements along the $[1\bar{1}0]_{CrN}$ ($\phi = 0°$) and $[1\bar{2}1]_{CrN}$ ($\phi = 90°$) in-plane directions on this CrN(111) layer show essentially identical $n$ and $k$ values (see Fig. S8 in the Supplemental Material), indicating that the CrN(111) layer exhibits largely isotropic optical properties. Together with the isotropic surface morphology of this CrN(111) layer, this finding supports the interpretation that the pronounced optical anisotropy of CrN(113) is related to its anisotropic surface morphology.

As an additional test of this interpretation, we investigated the in-plane optical anisotropy of the ScN(113) layer studied in Ref. 40. This layer was also grown on a semipolar AlN(11$\bar{2}$2) templates and possesses a surface morphology very similar to that observed for the CrN(113) layer (cf. Fig. 4 in this work and Fig. 1 in Ref. 40). As shown in Fig.S9 in the Supplemental Material, the optical anisotropy observed for this ScN(113) layer is smaller than for the CrN(113) layer shown above (Fig. 6), but is still on the order of 3–5%. This result provides further support to the proposed interpretation of the anisotropic surface morphology acting as a subwavelength grating.

The difference in magnitude of the optical anisotropy observed for the CrN(113) and ScN(113) layers can be understood by considering that while the surface morphology provides the geometrical structure inducing the optical anisotropy, it is the magnitude of the complex refractive index that determines how strongly light interacts with this structure. In particular, the imaginary part of the complex refractive index determines the optical penetration depth.

Figure S6 in the Supplemental Material shows the optical penetration depths of CrN and ScN across the investigated spectral range. This quantity is to be compared with the characteristic depth of the anisotropic surface modulation, which amounts to approximately 30 nm for both the CrN(113) and ScN(113) layers (cf. Fig. 4 in this work and Fig. 1 in Ref. 40). The depth of the surface modulation is comparable to the optical penetration depth in CrN for photon energies larger than 3 eV, consistent with the very large ($> 10\%$) optical anisotropy observed in the ultraviolet range. In contrast, the penetration depth in ScN exceeds the depth of the surface modulation in the entire spectral range (note that ScN becomes essentially transparent below its lowest direct band gap at approximately 2.2 eV[22]). For ScN, light of any wavelength at least partly probes the optically isotropic bulk. The difference in penetration depth thus provides a physically plausible explanation for why morphologically similar surface structures produce an optical anisotropy of different magnitude.

# 5 Summary and Conclusions

To summarize, we have demonstrated for the first time the presence of in-plane optical anisotropy in transition-metal nitrides, specifically in epitaxial CrN(113) layers grown on AlN(11$\bar{2}$2) templates by plasma-assisted molecular beam epitaxy. Structural analysis revealed a well-defined epitaxial relationship with a two-component lattice tilt arising from the strongly anisotropic lattice mismatch, thereby establishing the crystallographic basis for interpreting the optical response. By combining multi-azimuth spectroscopic ellipsometry with structural and surface analyses, we extracted the full in-plane dielectric response and quantified anisotropy levels of 10–12% for $n$ and 3–15% for $k$ across the 1.25–5 eV spectral range.

We have discussed the possible origins of this anisotropy in a nominally isotropic cubic crystal and find that its magnitude is incompatible with intrinsic mechanisms, while the essentially fully relaxed state of the layers rules out strain-induced contributions. The pronounced anisotropy is therefore attributed to the formation of an effective subwavelength grating by the highly anisotropic surface morphology. The weaker anisotropy observed for a morphologically similar ScN(113) layer further shows that the surface morphology alone does not determine the magnitude of the optical anisotropy; rather, its optical manifestation depends on the interplay between the characteristic dimensions of the surface structure and the complex refractive index of the material. In CrN(113), the characteristic depth of the surface modulation is comparable to the optical penetration depth, resulting in a strong contribution from the structured near-surface region. In ScN(113), by contrast, the substantially larger penetration depth reduces the relative contribution of the same type of surface modulation to the measured optical response.

# 6 Acknowledgments

We thank Carsten Stemmler for technical assistance with MBE#9, and Sebastian Henn for a critical reading of the manuscript. This work is funded by the Deutsche Forschungsgemeinschaft (DFG, Germany Research Foundation) within the Priority Programme SPP 2477: Nitrides4Future – Novel Materials and Device Concepts — project number 563156864.

## 7 Conflicts of Interest

The authors declare no conflicts of interest.

## 8 Data Availability Statement

The data that supports the findings of this study are available in the supplementary material of this article.

# Large in-plane optical anisotropy in epitaxial CrN(113) films: supplemental material

Duc V. Dinh,[*,†] Frank Peiris,[‡] and Oliver Brandt[†]

†*Paul-Drude-Institut für Festkörperelektronik, Leibniz-Institut im Forschungsverbund Berlin e.V., Hausvogteiplatz 5–7, 10117 Berlin, Germany*

‡*Department of Physics, Kenyon College, Gambier, 43022 Ohio, United States*

E-mail: duc.vandinh@pdi-berlin.de

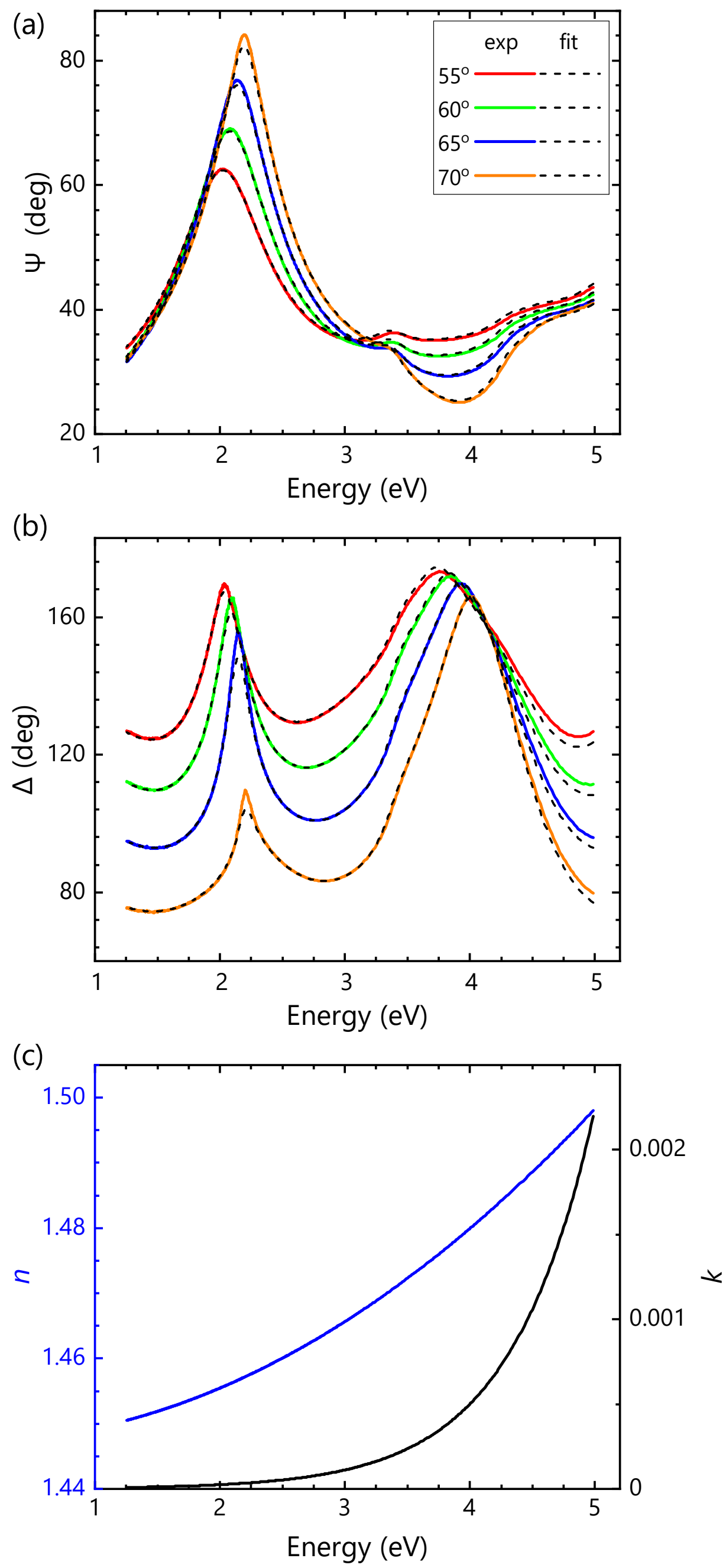

Fig. S1: (a) Ψ and (b) Δ spectra measured at different angles of incidence ($\Theta = 55 - 70°$) for a reference ≈ 125-nm-thick $SiO_2$/Si layer, together with their corresponding fits. (c) The extracted optical constants $n$ and $k$ of this layer.

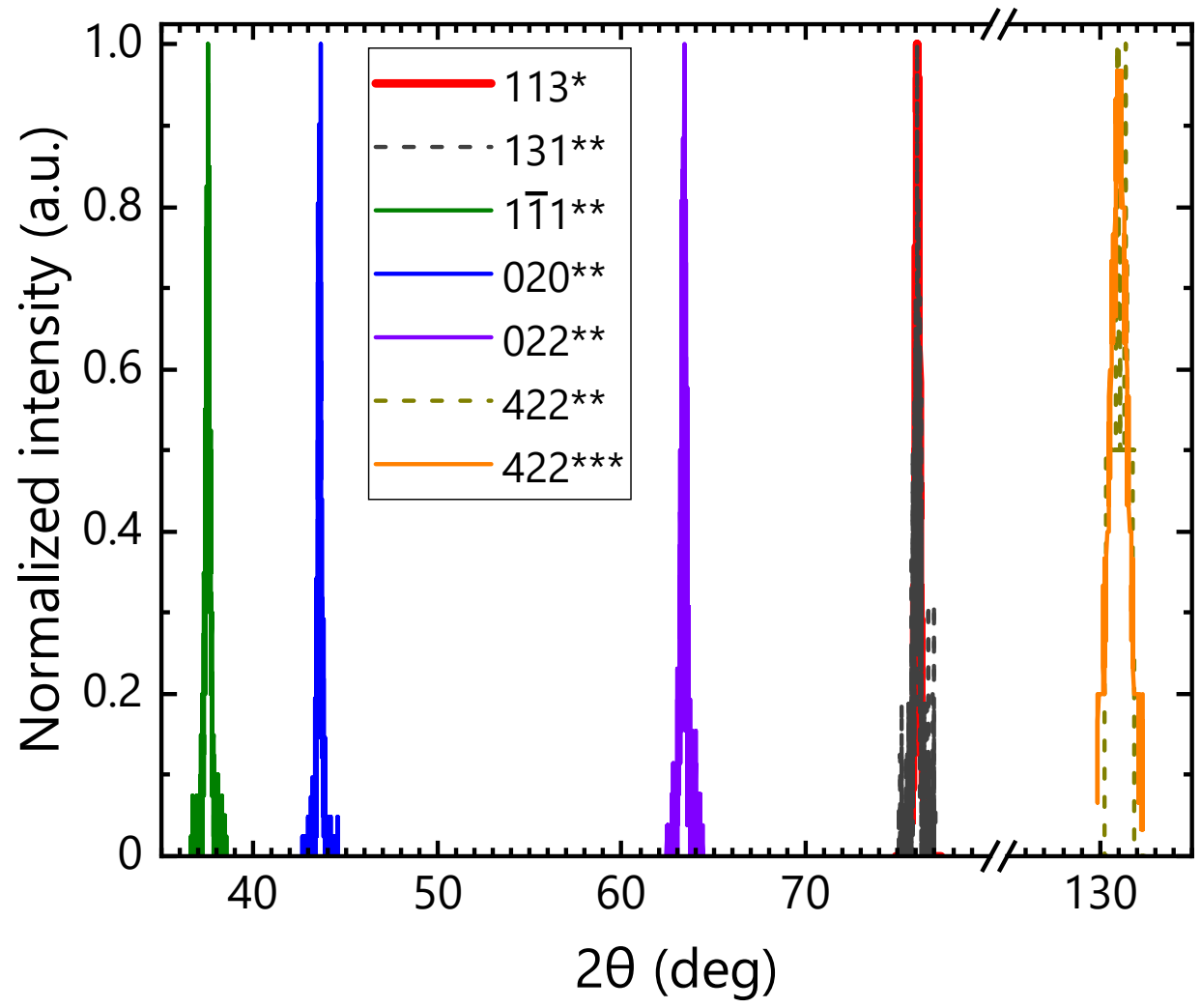

Fig. S2: X-ray diffraction symmetric (*), skew-symmetric (**), and asymmetric (***) reflections of a CrN(113) layer. The out-of-plane lattice constant along the $[113]_{\mathrm{CrN}}$ and the two in-plane lattice constants along the $[1\bar{1}0]_{\mathrm{CrN}}$ and $[33\bar{2}]_{\mathrm{CrN}}$ directions extracted from these reflections are 4.146 Å, 2.933 Å, and 9.727 Å, respectively. These results indicate that the layer is fully relaxed, with a residual strain magnitude of at most 0.03–0.06% along all axes.

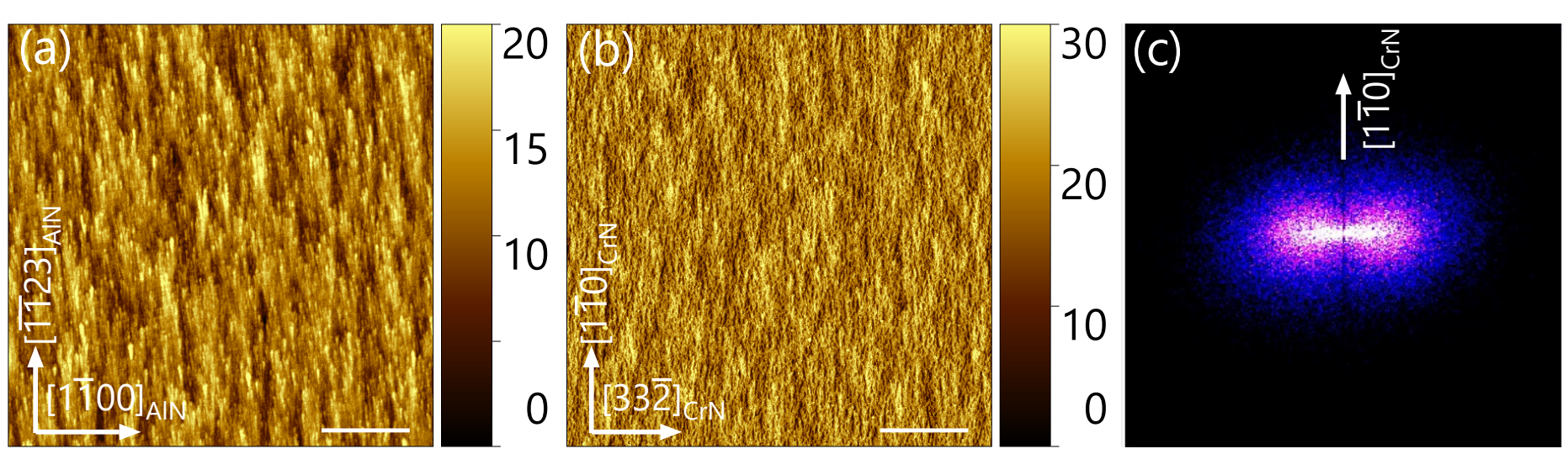

Fig. S3: $10 \times 10\,\mu\mathrm{m}^2$ atomic force topographs of (a) the AlN$(11\bar{2}2)$ template and (b) 200-nm-thick over-grown CrN(113) layer. The $z$ scale is in units of nm, and the scale bar is $2\,\mu$m. (c) Fast Fourier transform (FFT) pattern generated from (b) showing the uniaxial symmetry of this surface.

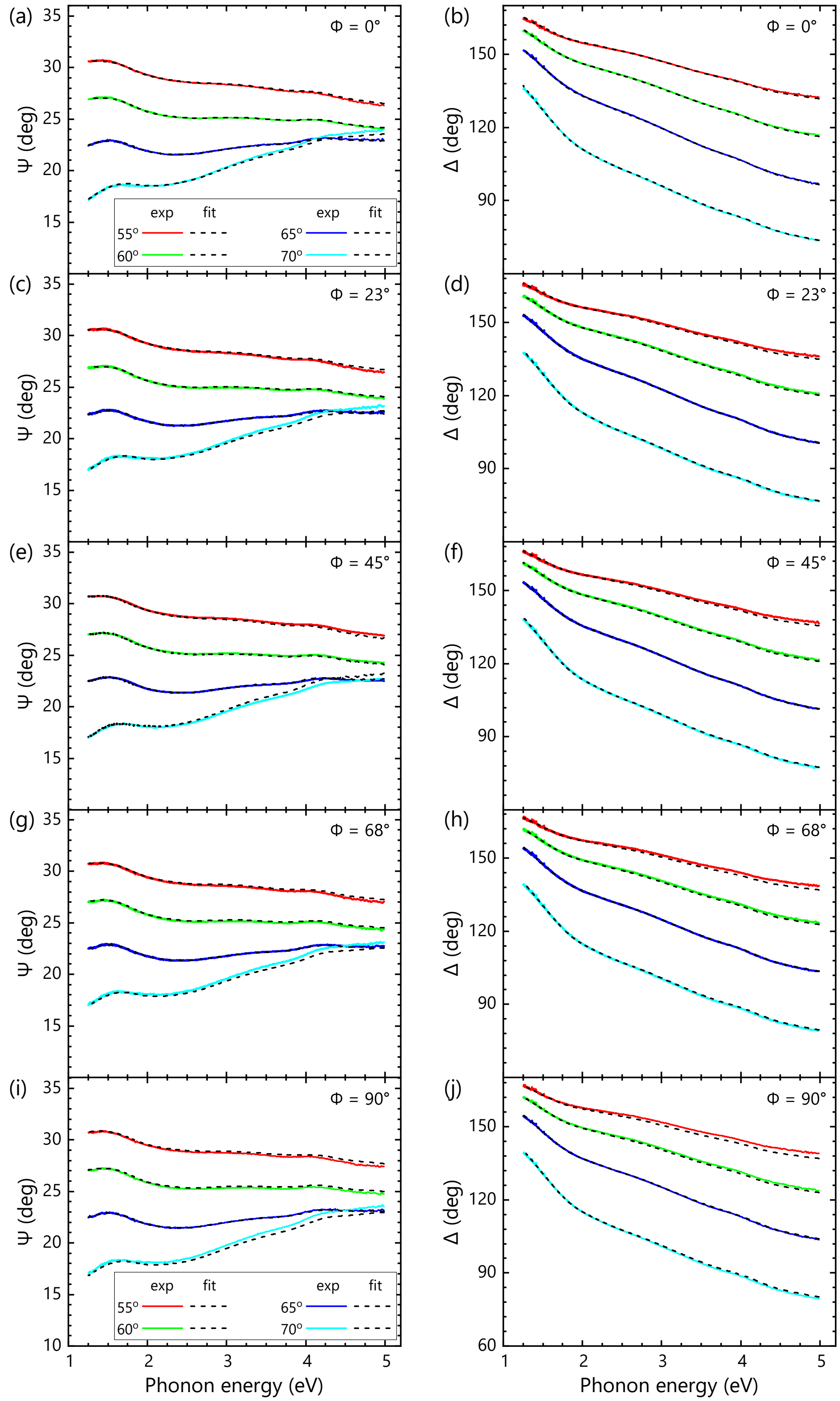

Fig. S4: Ψ (first column) and Δ (second column) spectra measured at different angles of incidence ($\Theta = 55 - 70°$) for the CrN(113) film, together with their corresponding fits (dashed lines). The five rows correspond to azimuthal angles: (a)–(b) $\phi = 0°$, (c)–(d) 23°, (e)–(f) 45°, (g)–(h) 68°, and (i)–(j) 90°, respectively. For simplicity, plot legends are shown only in (a) and (i); all other panels share the same line styles.

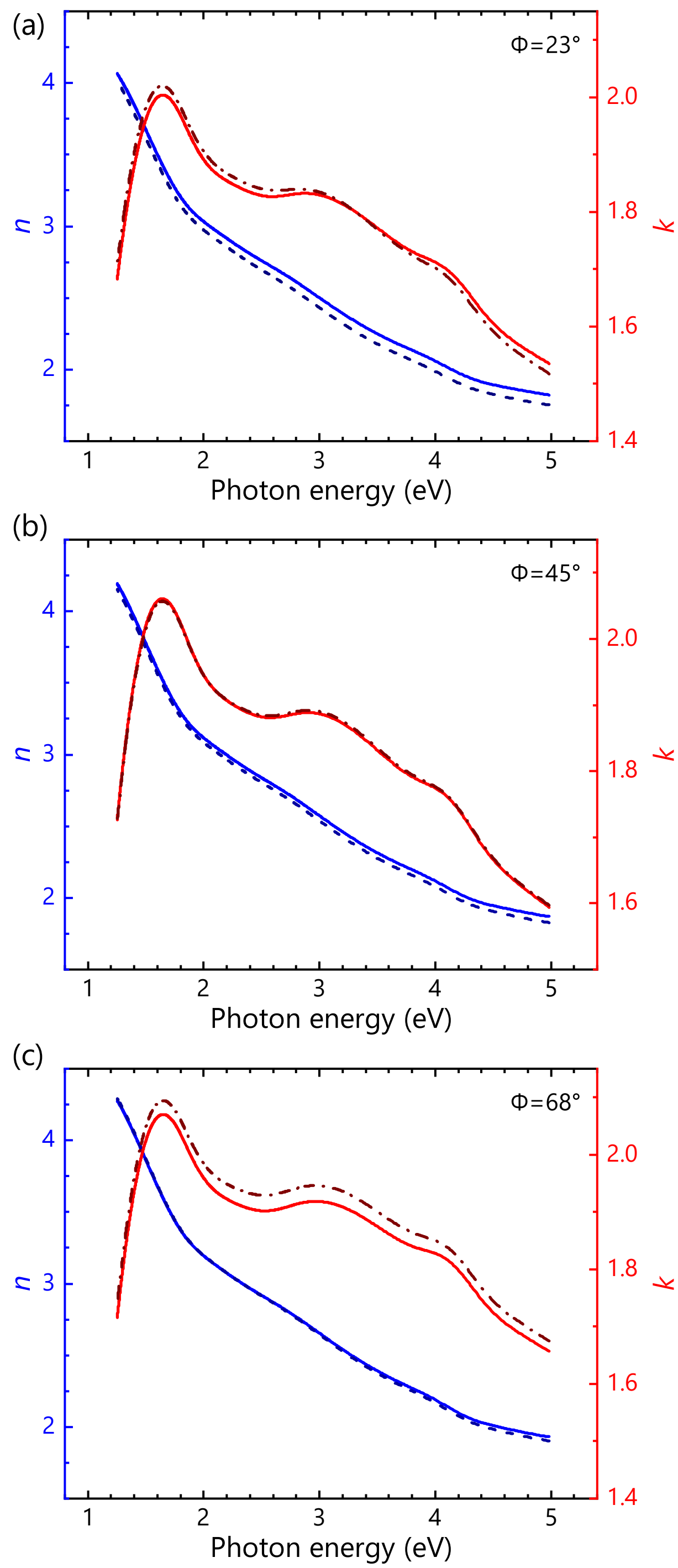

Fig. S5: Pseudo-optical constants $n$ and $k$ and calculated ones (dashed lines) of the CrN(113) film obtained at different azimuths: (a) $\phi$ = 23°, (b) $\phi$ = 45°, and (c) $\phi$ = 68°.

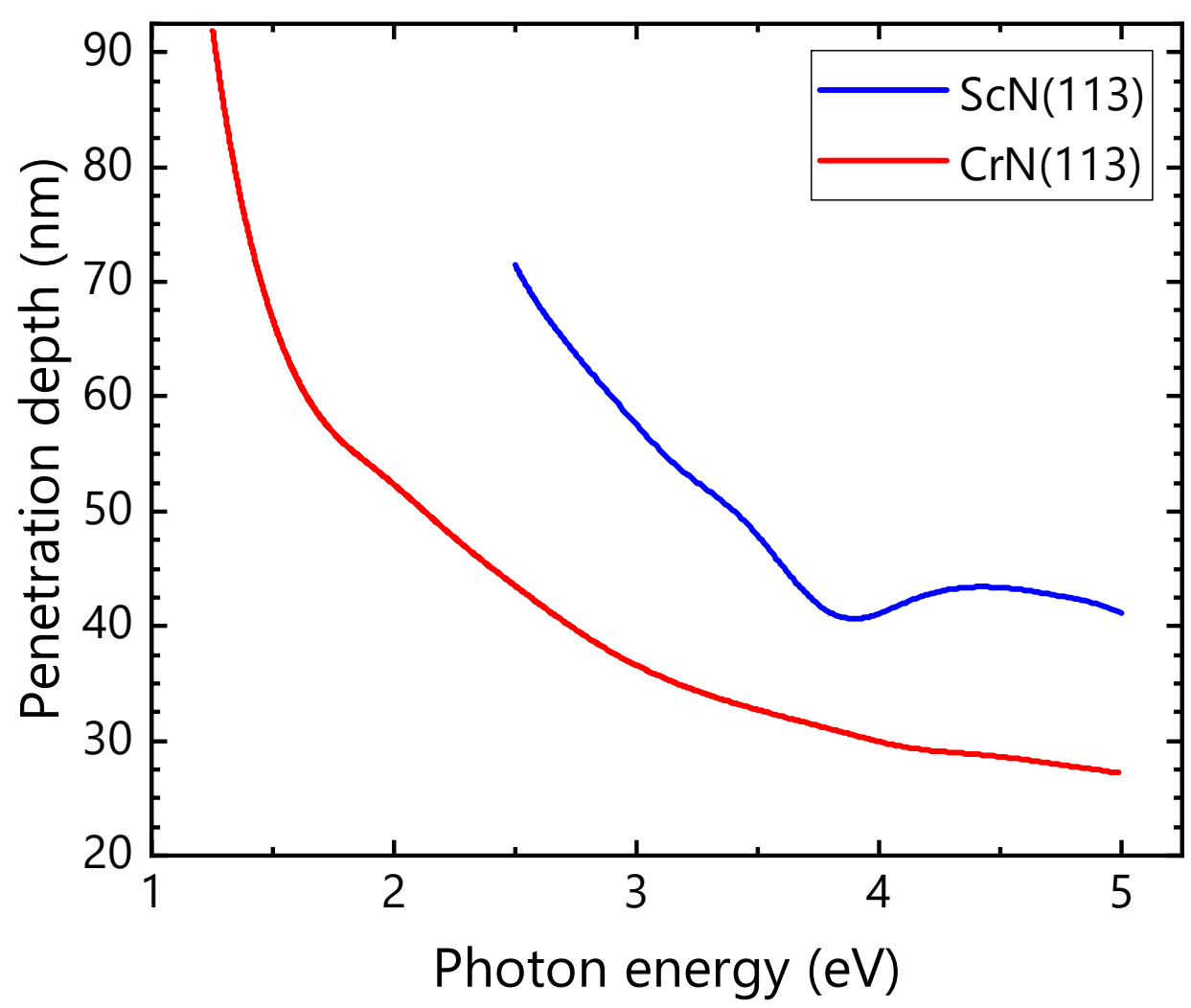

Fig. S6: Penetration depths of 200-nm-thick CrN(113) and ScN(113) layers for $\phi = 0°$ (along $[33\bar{2}]_{\mathrm{CrN, ScN}}$) calculated with for $\Theta = 0°$ (normal incidence). Note that the penetration depth of CrN(113) is even smaller than in the $\phi = 90°$ case due to the higher value of $k$ (see the main text for details). Also note that ScN becomes transparent below its lowest direct band gap at 2.2 eV.

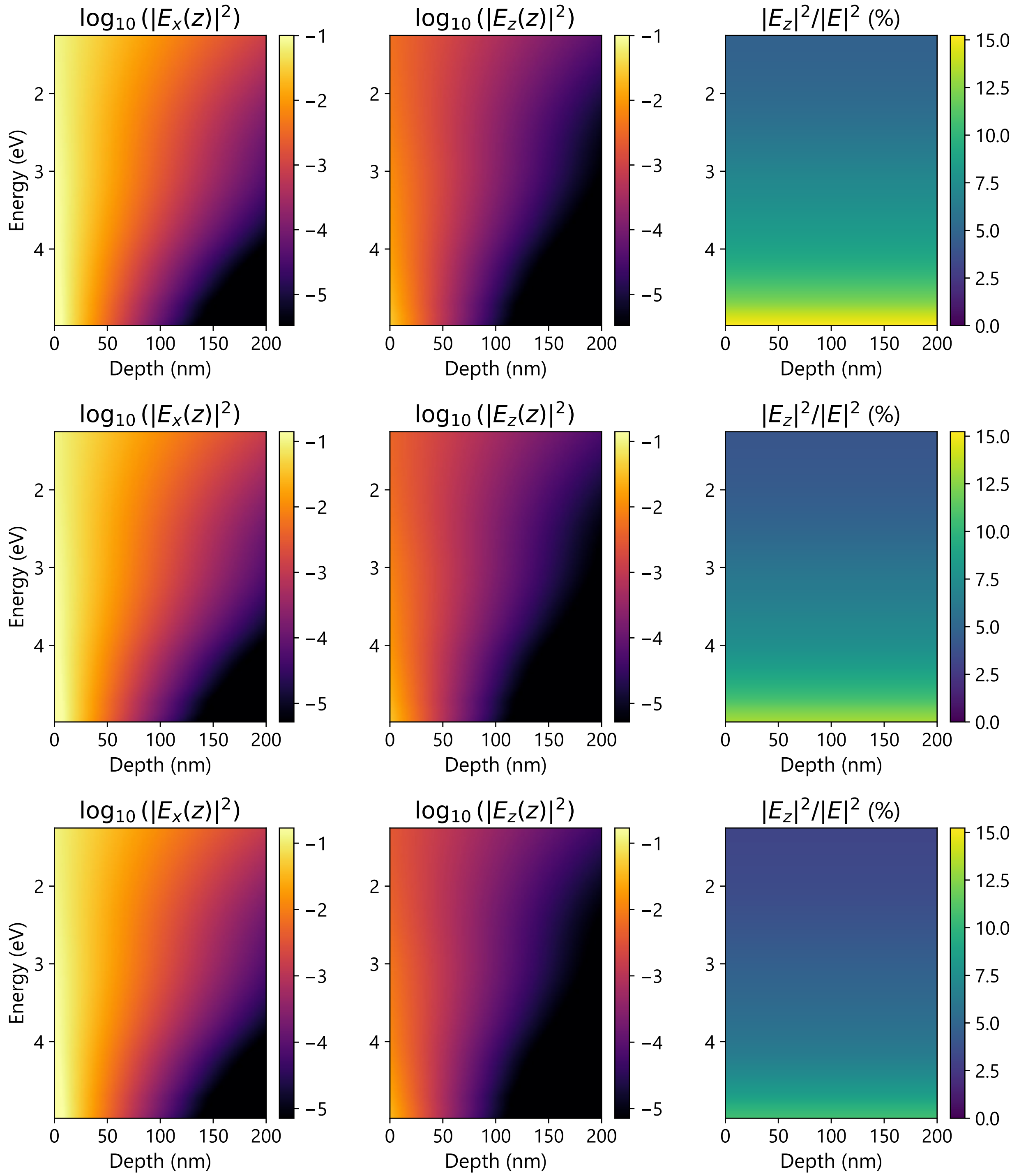

Fig. S7: In-plane $E_x(z)$ and out-of-plane $E_z(z)$ components, as well as the ratio $|E_z(z)|^2/|E|^2$ (where $|E|^2 = |E_x(z)|^2 + |E_z(z)|^2$), of the electric field calculated for CrN at angles of incidence of 50° (panel 1), 60° (panel 2), and 70° (panel 3). The calculations use the $n$ and $k$ values of CrN(113) obtained at an azimuthal angle $\phi = 0°$ (see main text). The results show that the out-of-plane field component remains small compared to the total field, leading to a weak contribution of the out-of-plane dielectric response $\varepsilon_{zz}$ to the ellipsometric spectra.

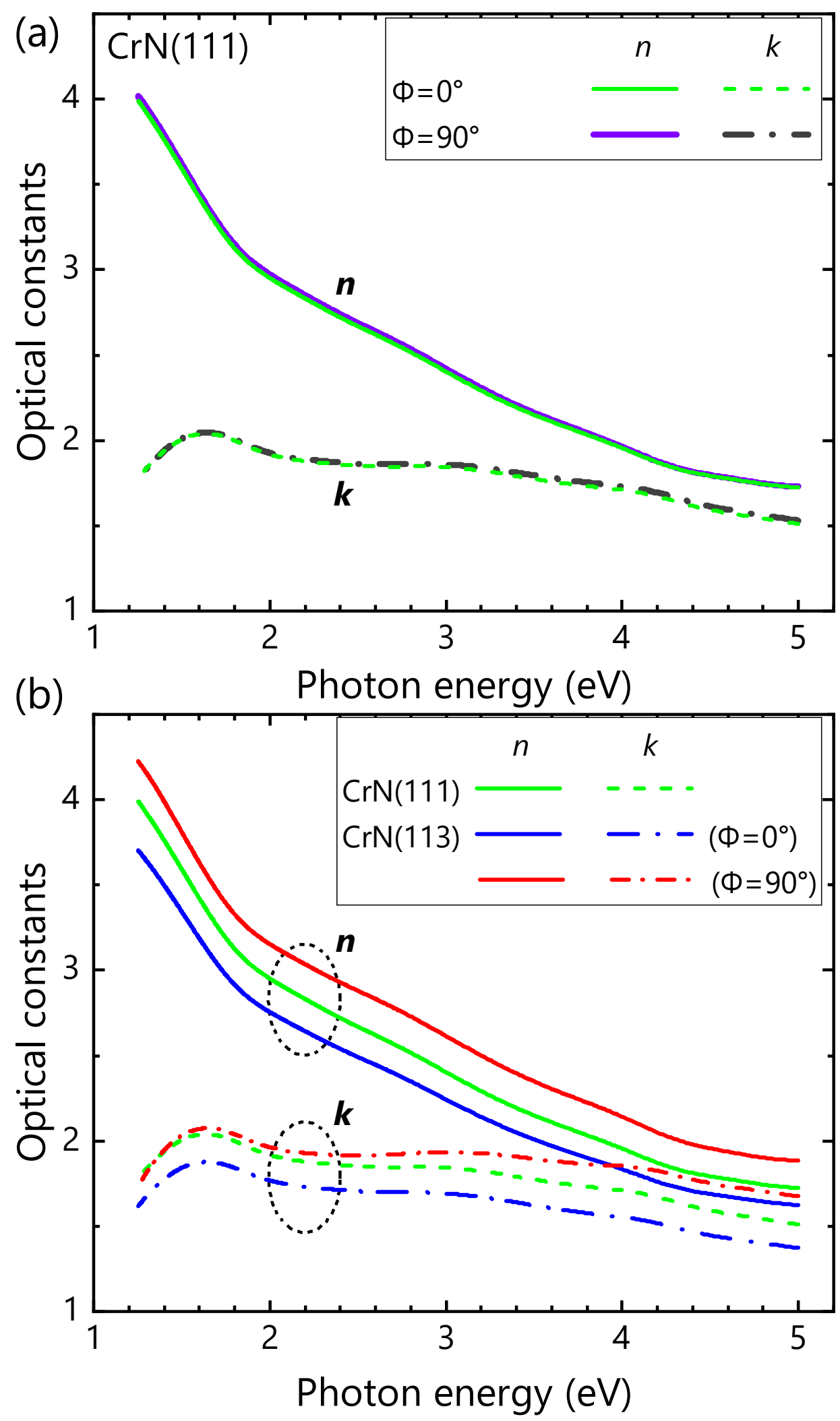

Fig. S8: (a) Pseudo-optical constants $n$ and $k$ of a CrN(111) reference sample grown on an AlN(0001) template, measured along two orthogonal in-plane directions: $[1\bar{1}0]$ ($\phi = 0°$) and $[1\bar{2}1]$ ($\phi = 90°$). (b) Comparison of the pseudo-optical constants $n$ and $k$ between the CrN(111) and CrN(113) samples, where the latter is measured along $[33\bar{2}]$ ($\phi = 0°$) and $[1\bar{1}0]$ ($\phi = 90°$). Interestingly, the pseudo-optical constants of the CrN(111) layer lie approximately between the extrema measured along the two principal in-plane directions of the CrN(113) layer for both $n$ and $k$.

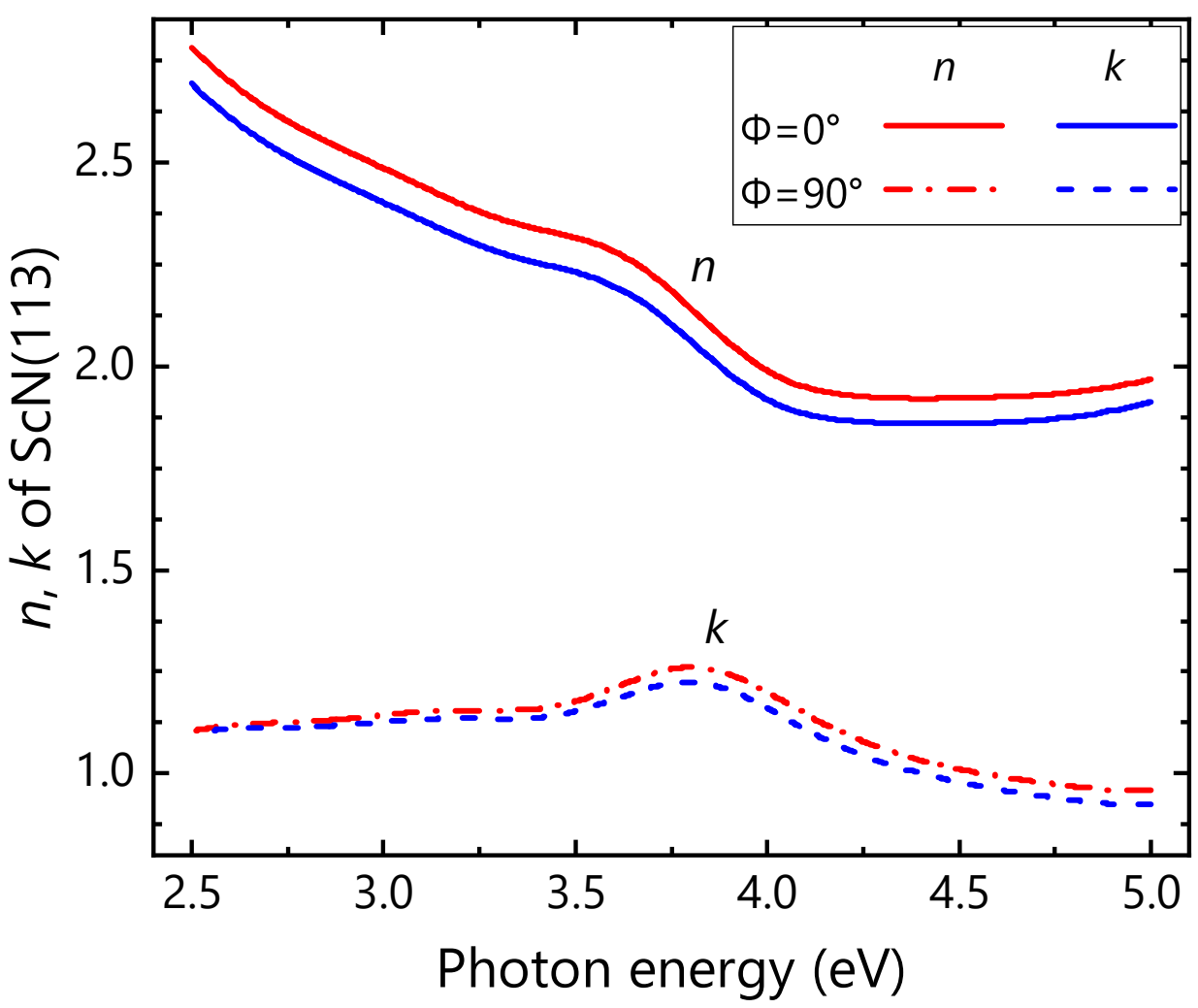

Fig. S9: Pseudo-optical constants $n$ and $k$ of a ScN(113) sample grown on an AlN(11$\bar{2}$2) template, measured along two orthogonal in-plane directions [33$\bar{2}$] ($\phi$ = 0°) and [1$\bar{1}$0] ($\phi$ = 90°). Note that we show only the absorptive region of ScN(113) to enable a fair comparison with the absorptive region of CrN(113) studied here.